# Preprint: Comparing Kinect2 based Balance Measurement Software to Wii Balance Board

Zhihan Lv, Vicente Penades, Sonia Blasco, Javier Chirivella, Pablo Gagliardo
FIVAN, Valencia, Spain
lvzhihan@gmail.com

## ABSTRACT
This is the preprint version of our paper on REHAB2015. A balance measurement software based on Kinect2 sensor is evaluated by comparing to Wii balance board in numerical analysis level, and further improved according to the consideration of BFP (Body fat percentage) values of the user. Several person with different body types are involved into the test. The algorithm is improved by comparing the body type of the user to the 'golden- standard' body type. The evaluation results of the optimized algorithm preliminarily prove the reliability of the software.

## Categories and Subject Descriptors
D.2.8 [Software Engineering]: Metrics—*complexity measures, performance measures*

## 1. INTRODUCTION
The center of pressure (CoP) is also known as center of mass (CoM). It is measured by different methods according to different technical levels in different age. The methods which are already applied in clinic include Balance Error Scoring System (BESS), 'gold-standard' laboratory-grade force plate (FP), wii blance board (WBB). Our research is to measure to what extend the kinect balance measurement (KBM) method can be accepted by clinical utilization. In kinesiology research community, currently, there are several kinds of measurements methods for comparison the 'gold-standard' laboratory-grade force plate (FP) to wii blance board (WBB) presented. There are some measurement results recorded in current literature. [42] compared the game scores provided by Wii to traditional balance measures. [7] wanted to see if traditional balance measures (like Center of Pressure velocity) were reliable on the wii balance board and valid compared to a forceplate, so they bypassed the games and simply collected the raw data. [1] performed a standard measurement uncertainty analysis to provide the repeatability and accuracy of the WBB force and CoP measurements. In summary, the data from a wii balance board is valid and reliable [16] [4] [7] but the game scores that are produced are not [42]. Moreover, WBB may be useful for low-resolution measurements, but should not be considered as a replacement for laboratory-grade force plates. Therefore, WBB is enough accurate to measure our @home rehabilitation application using kinect, but the customized software is expected to be developed.

In robotics research community, a series of researches have been already done about comparison kinect balance measurement (KBM) and Vicon balance measurement (VBM) to wii blance board (WBB) [17] [18] [19]. In the experiment, the WBB was placed on top of the FP to obtain measurement from both devices at the same time, which is set as the same as the method1 proposed in kinesiology research community. The person were instructed to stand on top of the FP and WBB and hold 40 static postures, each lasting 5s. [19] mentioned that the reason to measure the KBM is that the improper lighting, loose fitting clothes, and large objects which surround the subject can adversely influence the skeleton fitting. In our kinect2 based KBM method, we find age, height, body type of the person as well as the error in foot tip position measurement also affect the CoP measurement results slightly.

The purpose of our measurement is different from the previous measurements, since they just compared 'gold-standard' laboratory-grade force plate (FP) to a game controller wii balance board (WBB). It is undeniable that they proved that WBB can replace Balance Error Scoring System (BESS) which is used for assessing the balance as subjective method for past years. Even some productions are developed based their research, such as BtrackS [15] which is proved to reach the same level as WBB. But our KBM method is based on totally different measurement algorithm theory, implementation technology and suitable device context. We are using the range image data captured by kinect2 device in real-time, and further measure the CoP by our customized algorithm based on optical theory. The Wii balance board can cannot measure any information about the weight and CoP onside the device, while the kinect can get full information appeared in the camera view, but is not enable to retrieve the weight information since it is a non-contact sensor. Nonetheless, the same characteristics of both devices are that they have specialty or potential to measure CoP, but cannot measure density information by any direct or in-direct methods. Therefore, the robotics research community attracts our attention. The 'Center of mass calculation by kinect' [18] uses SESC as the core algorithm to solve the calculation of CoM. This work has been compared to some other work: 1. FP; 2. WBB; 3. Kinect; 4. Vicon (High quality camera); 5. Winter's method

## 2. SYSTEM

In our research, we will consider the previous proposed methods and results in both kinesiology research community and robotics research community, and design our particular measurement method according to our algorithm theory, device condition and clinical needs. The hardware devices we currently owned include: 1. WBB; 2. Several Kinect2. The algorithm that is similar with our algorithm is mentioned in literature [25] [21], which is so-called segmentation meth- ods, also known as kinematic methods. The problem that 'performed mainly on cadavers or in live, young and fit indi- viduals' is also mentioned [43] [13]. Some solutions are also discussed [34], which is 'should be adjusted for age, sex and fitness level'.

## 3. EVALUATION AND IMPROVING

The evaluation is conducted along with the improving process. The evaluation process includes several stages. During the evaluation, the subject should place the stance limb in the center of the Wii balance board and place his or her hands on the hips and eyes open. Next, he or she will reach as far as possible in the 3 reach directions with the contralat- eral limb. Reach distance was defined as the farthest point that an individual could touch without accepting weight and while maintaining balance through the return to a bilateral stance.

Four person with different body type were involved into the test. The purpose of this session is to find the relation between the body type with the errors of CoM calculation. Body segment inertial parameters (BSIPs) are important data in biomechanics. BSIPs have been measured by differ- ent methods, e.g. [3] [41] employs a whole body dual energy X-ray absorptiometry scan, [5] employs a motion capture system and two forceplates, [8] [9] employs a single kinect, [12] employs several digital cameras, [20] proposed a validation method to evaluate BSIPs.

Our balance measurement algorithm is using segmentation method, based on which, the weight of each segment's CoM position is the ratio of the segments to total body mass. BSIPs are the paramerhters for this purpose. It's already proved that BSIPs are depending on the age, gender and body type and some researches have modified the classic BSIPs based on this theory [45] [13] [14] [34] [36] [2] even concerning the effects of weight loss in obese individuals [32] and individuals of different morphology [10], as well as composite concerning. Methods relying on imagery have the drawback that the segmentation of body parts is complex, thus affecting significantly the BSIP values [20]. Indeed, the BSIP depend on the relative amount of bone, muscle and adipose tissue in a body segment which underlie structural and temporal changes when aging or caused by pathology or training [37] [24]. [35] [33] [23] have measured the extent to which errors in predicting body segment parameters could influence biomechanical analysis of human motion, predicted joint moment and joint kinetics.

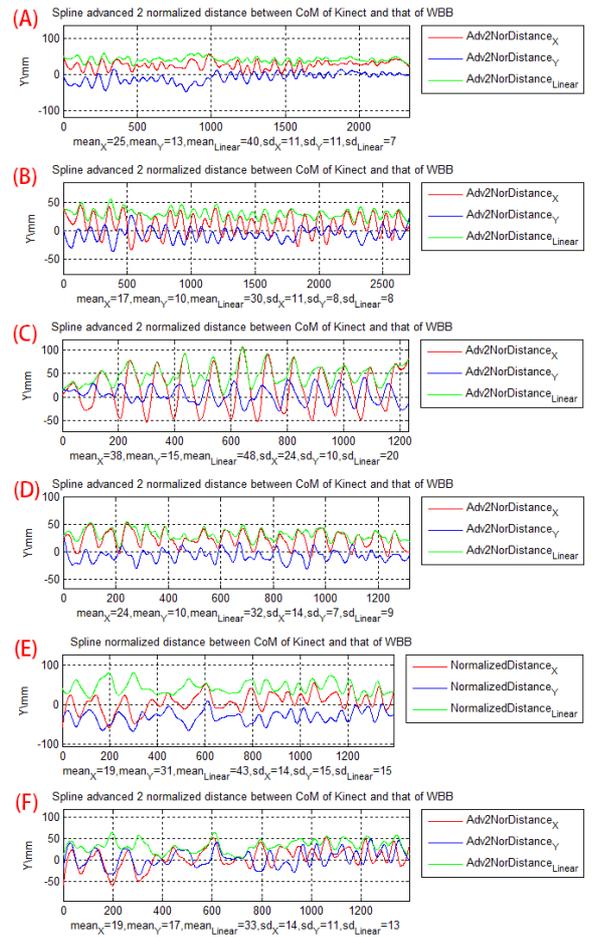

Figure 1: The evaluation results for four person.

This research is used to explorer the relationship between some metrics with the BSIPs and further improve the calculation of CoM. The most important factor is Body type, which is also known as body shape. Body shape is affected by body fat distribution, which is correlated to current levels of sex hormones. There are some existing metrics to evaluate the body type, such as BMI (Body Mass Index), BFP (Body fat percentage), BVI (Body Volume Index), WHR (Waist-hip ratio), WhtR (waist-to-height ratio). The series of im- ages of the results indicate that the normalized distances on X-axis between the CoM of Kinect and WBB haven't ob- vious regular patterns. While the normalized distances on Y-axis between the CoM of Kinect and WBB have the sim- ilar orbit that the normalized CoM of WBB is always after the normalized CoM of Kinect. It also reveals that the nor- malized distances on Y-axis between the CoM of Kinect and WBB depend on BFP. Because the order of the distances for the three subjects are $B(m = 67, sd = 11) > C(m = 48, sd = 18) > A(m = 13, sd = 11)$, which is the same as the order of $BFP\,(B(30.96) > C(23.39) > A(19.51))$. The
order of other metrics (i.e. BMI, WHR, WhtR) have differ-ent order.

BFP is also written as *BF* %, which is usually measured by physical device, such as underwater weighing, whole-body air displacement plethysmography, near-infrared interactance, dual energy X-ray absorptiometry. In our application scene, the physical device is not suitable. So we employes the formula derived from previous researches' statistic results. The calculation of BFP is as followed.

$$NEW\ BMI = 1.3 \times weight(kg)/height(m)^{2.5}\ [38]$$

*Adult body fat*% $=(1.20 \times BMI)+(0.23 \times Age)-(10.8 \times sex)-5.4$ where sex is 1 for males and 0 for females.

According to this supposed regular pattern, we proposed a new improvement plan, which considers the relative value of BFP of the player comparing to BFP of golden standard player (A). We will start from the derivation of a linear relation between BFP and the distance on Y-axis to represent the regular pattern. In contrast to the previous test for individual subject, we have considered the body types of three health people with different ages, height, weight this time. According to the BFP as well as the regular pattern indicated in the test results, the new refined formula for reducing error of distance between CoM of Kinect and WBB is proposed and could be abbreviated as:

$$0.0094 \times Y \times BFP^2 - 0.479 \times Y \times BFP + 5.732 \times Y \\ -19.47348 \times BFP^2 + 983.09884 \times BFP - 11762.064 \quad (1)$$

where $Y$ is the abbreviation of Y value of advanced normalized CoM by kinect, $BFP$ is the Body fat percentage value of the subject.

The evaluation of the new formula is conducted. Since the body type of A is supposed as golden standard in our research, so the error of distance remains the value (m=13, sd=11) after being refined by the new formula. This distance for B is vastly changed from (m=67, sd=11) to (m=10, sd=8). The result for C is also improved, even if it's not so effective as the improvement formula for individual in the previous research, but the error value of distance is acceptable. For the first session, the distance is changed from (m=32, sd=20) to (m=15, sd=10), comparing to the distance by formula for individual (m=13, sd=10). For the second session, the distance is changed from (m=48, sd=18) to (mean=10, sd=7) comparing to the distance by formula for individual (m=12, sd=9). The first session includes some extreme actions, so we can imagine that the measurement by Kinect is not so accurate as the second session. The distance curves for the three subjects are shown in figure 1(A)(B)(C)(D), where blue curves are error of distance on Y-axis.

Based on the optimization formula proposed before, we use another health people's data to test the optimized algorithm. The error is changed from (mean=31, sd=15)(in figure 1(E)) to (mean=17, sd=11)(in figure 1(F)). So far, the errors of all subjects are controlled into (mean=20).

## 4. CONCLUSION

More subjects will be involved into future tests to provide enough data for revealing the relation between body type (ie. BFP) with the CoM calculation. We will also consider the possibility to reuse the results of other researches about the relation between body type and BSIPs and further conclude the relation between body type and CoM calculation. Some novel technology will also be used to improve this research, e.g., Sensors [44], Virtual Rehabilitation [28] [27] [29] [30], Video Game [31] [6], Control [46], Database [40], Distributed Computing [39] [26] [22], Optimization Algorithm [11].

## Acknowledgment
The authors thank to LanPercept, a Marie Curie Initial Training Network funded through the 7th EU Framework Programme (316748).